\documentclass[10pt,aps,pra,twocolumn,groupedaddress]{revtex4-1}

\usepackage{amsmath}
\usepackage{graphicx}
\usepackage{color}
\usepackage{epstopdf}
\usepackage{hyperref}
\usepackage{layouts}
\usepackage{placeins}
\usepackage[danish,english]{babel}
\usepackage[utf8]{inputenc}

\newcommand{\JT}{\mathcal{A}}
\newcommand{\JS}{\mathcal{A}}

\newcommand{\vac}{\mathrm{vac}}
\newcommand{\bra}[1]{\left \langle #1 \right |}
\newcommand{\ket}[1]{\left | #1 \right \rangle}

\renewcommand{\H}{\hat{H}}

\newcommand{\ud}{\,\mathrm{d}}

\newcommand{\mean}[1]{ \left \langle #1 \right \rangle}

\renewcommand{\a}{\hat{a}}
\newcommand{\ad}{\hat{a}^\dagger}

\newcommand{\sinc}{\mathrm{sinc}}

\definecolor{col1}{RGB}{0,0,115}
\definecolor{col2}{RGB}{138,0,0}
\definecolor{col3}{RGB}{0,0,0}

\begin{document}


\title{Complete evolution equation for the joint amplitude in photon-pair generation}


\author{Jacob G. Koefoed}
\email[]{jgko@fotonik.dtu.dk}
\author{Karsten Rottwitt}
\affiliation{Department of Photonics Engineering, Technical University of Denmark, 2800 Kongens Lyngby, Denmark}

\date{\today}

\begin{abstract}
As four-wave-mixing-based photon-pair sources mature, accurate modelling of the photon-pair properties becomes important. Unlike spontaneous parametric down-conversion, four-wave mixing is accompanied by a number of parasitic effects such as nonlinear phase modulation. Currently, most modelling of photon-pair states are analytic in nature, which limits the number and type of effects that can be taken into account. In this work, we derive a complete, dual-pump evolution equation for the joint amplitude of photon pairs, wherein any desired effects can be included. We describe how to efficiently obtain numerical solution to this equation using a split-step approach. Lastly, we cover a few analytical solutions and compare two schemes for pure-photon generation under three different parasitic effects. We show how one scheme is highly sensitive to parasitic effects, while the other is very robust.
\end{abstract}


\maketitle

\section{Introduction}

Many quantum-optical technologies of the future, such as linear optical quantum computing~\cite{knill2001scheme}, rely on robust sources of highly indistinguishable photons. For many years, such single photons have been heralded from photon pairs produced by spontaneous parametric down-conversion (SPDC) in nonlinear crystals~\cite{kwiat1995new}. More recently there has been an increased interest in photon-pair generation through spontaneous four-wave mixing (SpFWM). Compared to SPDC there are several advantages to using SpFWM such as in-fiber generation~\cite{fiorentino2002all, li2005optical, sharping2004quantum, rarity2005photonic} and additional flexibility from multiple pumps, but the main one is perhaps ease of integration into established integrated platforms such as silicon~\cite{sharping2006generation, xiong2011slow}. This had already led to large-scale systems based on SpFWM sources~\cite{wang2018multidimensional}.

Many proposed quantum-optical technologies rely on two-photon interference~\cite{hong1987measurement}, which requires that the photons are indistinguishable. However, when detecting a photon-pair member during the heralding process, the remaining photon is projected into an impure quantum state, unless the two photons are completely uncorrelated in time and frequency~\cite{grice1997spectral, uren2005}. One simple solution is to employ narrow spectral filters~\cite{branczyk2010optimized}, but at the cost of increased system loss and heralding efficiency~\cite{christ2012limits}. To avoid this, a multitude of schemes for generating pure photons without spectral filtering in crystals~\cite{mosley2008heralded}, fibers~\cite{halder2009nonclassical, cohen2009tailored, clark2011intrinsically, soller2011high}, ring resonators~\cite{vernon2017truly, christensen2018engineering} and more have been proposed.

The usual way to determine the correlations in the biphoton state is to find approximate analytical solution to the quantum equations. However, for real systems where additional parasitic effects are included, such solutions may not exist. Many parasitic effects have been shown to degrade photon purity, such as group-velocity dispersion (GVD)~\cite{koefoed2019effects}, nonlinear phase modulation (NPM)~\cite{bell2015effects, christensen2016temporally} or dispersion fluctuations (DFs) due to longitudinal variations in waveguide properties such as refractive index or cross-sectional geometry~\cite{cui2012spectral, francis2016characterisation, koefoed2017spectrally}. If multiple of these or other detrimental effects are consequential for the biphoton state, a more generally applicable approach, such as a general numerical solver, is needed.

In this work, we derive, from the Heisenberg equations of the field operators, a general evolution equation for the joint amplitude of photon pairs generated by SpFWM and include the effects of GVD, NPM and DFs. This is a fully Schrödinger description and no further reference to any quantum operators is needed in order to fully describe the photon-pair state, including the photon-photon spectral and temporal correlations. We describe a numerical split-step algorithm for solving the propagation equation. We then show how to obtain analytical solutions to this equation when each effect is included individually and demonstrate the effects and their consequences individually in an example waveguide.

\section{Derivation of the evolution equation}
\subsection{The interaction picture}

A common starting point for the analysis of the two-photon state is the coupled Heisenberg equations for the signal and idler field operators $\a_s$ and $\a_i$. We use normalizations such that the equal-position commutator takes the form $[\a_j(z,t), \ad_k(z,t')] = \delta_{jk}\delta(t-t')$. In this case, the field operators satisfy the Heisenberg equations~\cite{bell2015effects}:
\begin{subequations}
\label{eq:HB}
\begin{align}
\partial_z \a_s &= i\frac{\Delta\beta_0(z)}{2}\a_s-\beta_{1s} \partial_t \a_s -  \frac{i}{2}\beta_{2s}\partial_t^2 \a_s \notag \\
&\quad+ 2i\gamma_{sp} |A_p|^2\a_s + 2i\gamma_{sq} |A_q|^2\a_s\notag + i\gamma A_p A_q \ad_i,  \\
\partial_z \a_i &= i\frac{\Delta\beta_0(z)}{2}\a_i -\beta_{1i} \partial_t \a_i -  \frac{i}{2}\beta_{2i}\partial_t^2 \a_i \notag \\
&\quad+ 2i\gamma_{ip} |A_p|^2\a_i + 2i\gamma_{iq} |A_q|^2\a_i\notag + i\gamma A_p A_q \ad_s.
\end{align}
\end{subequations}
Here, $A_j$ with $j = p,q$ describes slowly-varying classical pump-field envelopes, and $\Delta\beta_0(z) =   \beta_{0s}(z) + \beta_{0i}(z) - \beta_{0p}(z) - \beta_{0q}(z)$ is the waveguide-position dependent phase mismatch. The dispersion parameters $\beta_{nj}$, $n = 0,1,2$, $j = s,i,p,q$, describes the $n$th derivative of the propagation constant of field $j$ with respect to frequency at the central frequency of the field. The nonlinear parameters $\gamma_{jk}$ describes the nonlinear interaction strength between fields $j$ and $k$, while $\gamma$ is the four-wave mixing nonlinearity. For identical waveguide modes and copolarized fields, all the nonlinear parameters are identical.

While these equations fully describe the quantum evolution of the fields $\a_s$ and $\a_i$ given the pumps $A_p$ and $A_q$, they are still operator equations making them difficult to handle numerically. In the literature, these equations are either solved directly~\cite{lin2007photon} (when exact solutions are available), using Green functions to obtain an input-output relation for the quantum fields \cite{vernon2015spontaneous, koefoed2019effects, christensen2016temporally} or in the interaction picture \cite{bell2015effects, koefoed2017spectrally}. Instead, we seek to derive an evolution equation for the joint spectral wavefunction of the photon pair.

We transition to the interaction picture by splitting the total system Hamiltonian into two parts, $\H = \H_0 + \H_1$, where $\H_1$ contains the four-wave-mixing interaction and $\H_0$ governs dispersion and nonlinear phase modulation. In the interaction picture, the state is then governed by
\begin{equation}
\H_\mathrm{int}(z) =  \gamma \int \ud t  A_p(z,t) A_q(z,t) \ad_s(z,t) \ad_i(z,t) + \mathrm{H.c.}
\end{equation}
which is identical to $\H_1$, but with the Schrödinger operators replaced by Heisenberg operators evolving under $\H_0$. Under this interaction, the system state $\ket{\psi}$ evolves according to
\begin{equation}
\frac{d}{dz}\ket{\psi} = i \H_\mathrm{int} \ket{\psi}.
\end{equation}
When analyzing the spectral and temporal properties of photon pairs, it is convenient to express the biphoton part of the state as
\begin{equation}
\ket{\psi_\mathrm{bi}(z)} = \iint \ud t_s \ud t_i \JT(z,t_s,t_i) \ad_s(z,t)\ad_i(z,t) \ket{\vac},
\end{equation}
where the joint temporal amplitude (JTA) $\JT(z,t_s,t_i)$, which is simply a joint wavefunction for the photons in the time domain, contains all information on the temporal components of the photons and their correlations. By this definition, the JTA can easily be extracted from the total system state:
\begin{equation}
\JT(z,t_s,t_i) = \bra{\vac}\a_s(z,t) \a_i(z,t) \ket{\psi}.
\end{equation}
To discover an evolution equation for the JTA, we take the spatial derivative of this expression
\begin{align}
\frac{\partial \JT(z,t_s,t_i)}{\partial z} &= i\bra{\vac}\frac{\partial}{\partial z} [\a_s(z,t_s) \a_i(z,t_i)] \ket{\psi} \notag \\ 
&\quad + \bra{\vac}\a_s(z,t) \a_i(z,t) \frac{\partial}{\partial z}\ket{\psi}.
\end{align}
The first term covers all effects included in the field operator evolution (dispersion, nonlinear phase modulation) while the second term is FWM. The first term is straight-forward to evaluate using the Heinsenberg equations \eqref{eq:HB} (without the FWM term). The second term is
\begin{align*}
\bra{\vac}\a_s(z,t_s) &\a_i(z,t_i) \frac{\partial}{\partial z}\ket{\psi} \\
& = \bra{\vac}\a_s(z,t_s) \a_i(z,t_i)\H_\mathrm{int}\ket{\psi} \\
&= i\gamma\int \ud t A_p(z,t) A_q(z,t) \\
&\quad\times \bra{\vac}\a_s(z,t_s) \a_i(z,t_i)\ad_s(z,t) \ad_i(z,t)\ket{\psi} \\
&= i\gamma \delta(t_s - t_i) A_p(z,t_s) A_q(z,t_s) \mean{\vac | \psi},
\end{align*}
where the last step used the field commutators to move the field operators. The inner product $\mean{\vac | \psi}$ is non-trivial to evaluate, but fortunately for photon-pair generation it is always close to 1, which is consistent with the perturbative approach usually taken when calculating photon-pair states. This approximation leads to the evolution equation for the JTA:
\begin{widetext}
\begin{align}
\label{evolution_equation}
\frac{\partial \mathcal{A}(z,t_\mathrm{s},t_\mathrm{i})}{\partial z} &= i\gamma A_\mathrm{p}(z,t_\mathrm{s}) A_\mathrm{q}(z,t_\mathrm{s}) \delta(t_\mathrm{s}-t_\mathrm{i}) + 
i\left [ \Delta\beta_0(z) + i\beta_{1\mathrm{s}} \frac{\partial}{\partial t_\mathrm{s}} + i\beta_{1\mathrm{i}} \frac{\partial}{\partial t_\mathrm{i}} - \frac{1}{2}\beta_{2\mathrm{s}}\frac{\partial^2}{\partial t_\mathrm{s}^2} 
 - \frac{1}{2}\beta_{2\mathrm{i}}\frac{\partial^2}{\partial t_\mathrm{i}^2} \right ] \mathcal{A}(z,t_\mathrm{s},t_\mathrm{i}) \notag\\
&+2i \left [\gamma_{sp}|A_\mathrm{p}(z,t_\mathrm{s})|^2 + \gamma_{sq}|A_\mathrm{q}(z,t_\mathrm{s})|^2 + \gamma_{ip}|A_\mathrm{p}(z,t_\mathrm{i})|^2 + \gamma_{iq}|A_\mathrm{q}(z,t_\mathrm{i})|^2  \right ] \mathcal{A}(z,t_\mathrm{s},t_\mathrm{i}).
\end{align}
\end{widetext}
This evolution equation contains three effects that are not usually considered in the context of FWM photon-pair generation. The first, which we call dispersion fluctuations (DFs), is longitudinal variation in the phase-matching condition through $\Delta\beta(z)$.
In realistic systems this is an important effect limiting single-photon purity~\cite{cui2012spectral, francis2016characterisation, koefoed2017spectrally}. In this work, we employ a simple model for DFs where the phase-matching frequency is varying through a Langevin process. To simulate a real system, the DFs should be linked to some underlying physical fluctuations such as waveguide geometry or index profile. The important fluctuations can be different in e.g. photonic-crystal fibers~\cite{francis2016characterisation,cui2012spectral} and step-index fibers~\cite{koefoed2017effects}. Longitudinal variations of other parameters could likewise be included, but are rarely significant~\cite{koefoed2017spectrally}. 
The second effect included is higher-order dispersion (HOD), where we here only include group-velocity dispersion (GVD) through the parameters $\beta_{2j}$. This effect can be significant or negligible, depending on the scheme considered~\cite{bell2015effects, koefoed2019effects}. The third effect is nonlinear phase modulation (NPM), included through the four last terms in the equation. Like GVD, this effect is sometimes consequential~\cite{bell2015effects} and sometimes not~\cite{christensen2016temporally, sinclair2016effect}. 

In fiber-based systems we expect these three effects to be the most consequential. Other waveguide platforms could have other parasitic effects than the ones included in this work. For example, accurate modelling of silicon waveguides could require the inclusion of two-photon absorption or free-carrier absorption. Additional effects can be included by following the procedure outlined here, starting from the Heisenberg equation for the field operators. Lastly, the delta-function in the FWM term originates from the near-instantaneous nature of the electronic nonlinear response, but it can be modified to account for a finite response time, for example when the Raman effect is considered~\cite{koefoed2017effects}.

\section{Split-step scheme for obtaining numerical solutions}

The evolution equation \eqref{evolution_equation} only allows analytical solution in special cases. However, in real systems many parasitic effects need to be included in the model. This requires a numerical routine that can efficiently generate solutions for any realistic system. Due to its similarity to the nonlinear Schrödinger equation, the evolution equation for the JTA can be solved by a similar split-step approach. Such an approach has previously been used succesfully in the degenerate pump case \cite{bell2015effects, koefoed2017effects}. Here, we outline the procedure for arbitrary non-degenerate pumps and discuss how to apply the steps corresponding to different effects.

\begin{figure}[!htb]
    \centering
    \def\svgwidth{0.95\linewidth}
    
    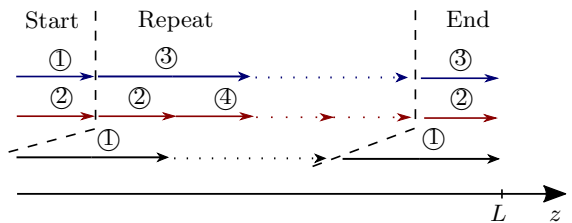
    \caption{The initializing, repeating and finalization parts of the split-step algorithm. Each part is bounded by the dashed line and the order of steps in each part is indicated with numbers. The spontaneous scattering effects ($\mathcal{S}$) are indicated in blue, the linear effects ($\mathcal{L}$) are in red while the nonlinear effects ($\mathcal{N}$) are in black. Small arrows represent a half-step of $\Delta z /2$ while long arrows represent a full step of $\Delta z$.}
    \label{fig:split_steps}
\end{figure}

We first define the operators
\begin{subequations}
\begin{align}
\mathcal{N} &= 2i \big [\gamma_{sp}|A_\mathrm{p}(z,t_\mathrm{s})|^2 + \gamma_{sq}|A_\mathrm{q}(z,t_\mathrm{s})|^2 \notag \\
&\quad + \gamma_{ip}|A_\mathrm{p}(z,t_\mathrm{i})|^2 + \gamma_{iq}|A_\mathrm{q}(z,t_\mathrm{i})|^2  \big ], \\
\mathcal{L} &= i\Bigg [ \Delta\beta_0(z) + i\beta_{1\mathrm{s}} \frac{\partial}{\partial t_\mathrm{s}} + i\beta_{1\mathrm{i}} \frac{\partial}{\partial t_\mathrm{i}} \notag\\
&\quad- \frac{1}{2}\beta_{2\mathrm{s}}\frac{\partial^2}{\partial t_\mathrm{s}^2} 
 - \frac{1}{2}\beta_{2\mathrm{i}}\frac{\partial^2}{\partial t_\mathrm{i}^2} \Bigg ], \\
\mathcal{S} &= i\gamma A_\mathrm{p}(t_\mathrm{s}) A_\mathrm{q}(t_\mathrm{s}) \delta(t_\mathrm{s}-t_\mathrm{i}).
\end{align}
\end{subequations}
Using these operators, we can write the evolution equation \eqref{evolution_equation} in the simple form
\begin{equation}
\frac{\partial \JT}{\partial z} = (\mathcal{N} + \mathcal{L})\JT + \mathcal{S}.
\end{equation}
This first-order partial differential equation has the formal solution
\begin{align}
&\JT(z+\Delta z) = \exp\left (\int_z^{z+\Delta z} \ud z' [\mathcal{L} + \mathcal{N}] \right ) \\
&\times \left [\JT(z) + \int_z^{z+\Delta z} \ud z' \exp\left ( - \int_z^{z'} \ud z'' [\mathcal{L} + \mathcal{N}] \right ) \mathcal{S} \right ]. \notag
\end{align}
Approximating the second integral with the trapezoidal rule $\int_z^{z+\Delta z} \ud z' \, f(z') = [f(z) + f(z+\Delta z)]\Delta z/2  + \mathcal{O}(\Delta z^3)$ yields
\begin{align}
\JT(z+\Delta z) &= \left [\JT(z)+ \frac{\Delta z}{2}\mathcal{S}(z)\right ]\notag \\
&\quad\times\exp\left (\int_z^{z+\Delta z} \ud z' [\mathcal{L} + \mathcal{N}] \right )\notag \\
&+ \frac{\Delta z}{2}\mathcal{S}(z+\Delta z) + \mathcal{O}(\Delta z^3).
\label{eq:steps}
\end{align}	
From the regular symmetrized split-step schemes, we also know that the application of the linear and nonlinear steps has a local error $\mathcal{O}(\Delta z^3)$ if half a linear step is applied, followed by a full nonlinear step and ended with another half linear step~\cite{Agrawal2006}. Using this, the total local error is $\mathcal{O}(\Delta z^3)$. This is achieved by the order of steps illustrated in Fig. \ref{fig:split_steps} and in accordance with Eq. \eqref{eq:steps} starts with a half-step of both spontaneous scattering and linear effects. This is followed by as many repetitions as needed of: A full nonlinear step, half a linear step, a full spontaneous scattering step and another half linear step. To bring all effects to the full propagation distance, the process is finalized by a full nonlinear step, a half linear step and a half spontaneous scattering step.
If the algorithm is performed in this way instead of a more straight-forward application of steps, it is simpler and more efficient to apply the spontaneous scattering step in the frequency domain, in which case it takes the form of a convolution of the pump spectra. In the frequency domain with the Fourier transform convention $f(\omega) = \int \ud t f(t) \exp(i\omega t)$, the spontaneous scattering and linear effects take the form
\begin{align}
\tilde{\mathcal{S}}(z,\omega_s, \omega_i) &= \frac{i\gamma}{2\pi} \int \ud \omega A_p(\omega) A_q(z,\omega_s + \omega_i -\omega), \\
\tilde{\mathcal{L}}(z,\omega_s,\omega_i) &= i \Bigg[ \Delta\beta_0(z) + \beta_{1\mathrm{s}} \omega_s + \beta_{1\mathrm{i}}\omega_i\notag \\
&\quad + \frac{1}{2}\beta_{2\mathrm{s}}\omega_s^2 + \frac{1}{2}\beta_{2\mathrm{i}}\omega_i^2 \Bigg],
\end{align}
where tilde denotes the Fourier transform. By approximating the integrals in Eq. \eqref{eq:steps} with the trapezoidal method, the solutions for a full step of each effect are
\begin{subequations}
\begin{align}
 \mathcal{A}_\mathcal{N}(z+\Delta z)&= \mathcal{A}(z) \exp\left \{\left [\mathcal{N}(z) + \mathcal{N}(z+\Delta z) \right ] \frac{\Delta z}{2} \right \},\\
\tilde{\mathcal{A}}_\mathcal{L}(z+\Delta z) &= \tilde{\mathcal{A}}(z) \exp\left \{\left [\tilde{\mathcal{L}}(z) + \tilde{\mathcal{L}}(z+\Delta z) \right ] \frac{\Delta z}{2} \right \}, \\
 \tilde{\mathcal{A}}_\mathcal{S}(z+\Delta z)&= \tilde{\mathcal{A}}(z) + \tilde{\mathcal{S}}(z)\Delta z.
\end{align}
\end{subequations}
Another possibility is to interchange the linear and nonlinear steps and apply the spontaneous step in the time domain. However, in this case, the temporal delta function must implemented carefully to avoid numerical artefacts. Lastly we note that, in many cases it may be sufficient to reduce the step-size $\Delta z$ to obtain the required precision and not worry about the optimal ordering of steps.


\FloatBarrier
\section{Special case analytical solutions}

In this section we provide a few special-case solutions to the general evolution equation \eqref{evolution_equation}.

\subsubsection{Solution with dispersion fluctuations and nonlinear phase modulation}

To eliminate the single temporal derivatives in the evolution equation \eqref{evolution_equation}, we employ the transformations
\begin{subequations}
\begin{align}
z' &= z, \\
t_s' &= t_s + \beta_{1s} (L-z), \\
t_i' &= t_i + \beta_{1i} (L-z), 
\end{align}
\end{subequations}
where $L$ is the waveguide length. This transforms the evolution equation into (relabelling the primed variables into non-primed variables)
\begin{align}
\frac{\partial \mathcal{A}}{\partial z} &= i\gamma A_\mathrm{p}(z,t_\mathrm{s} - \beta_{1s}(L-z)) A_\mathrm{q}(z,t_\mathrm{s} - \beta_{1s}(L-z))\notag\\
&\quad\times \delta(t_\mathrm{s}-t_\mathrm{i} - (\beta_{1s} - \beta_{1i}) (L-z)) \notag \\
&\quad + i\left [ \Delta\beta_0(z) - \frac{1}{2}\beta_{2\mathrm{s}}\frac{\partial^2}{\partial t_\mathrm{s}^2} 
 - \frac{1}{2}\beta_{2\mathrm{i}}\frac{\partial^2}{\partial t_\mathrm{i}^2} \right ] \mathcal{A}(z,t_\mathrm{s},t_\mathrm{i}) \notag\\
&+2i \Big [\gamma_{sp}|A_\mathrm{p}(z,t_\mathrm{s} - \beta_{1s}(L-z))|^2 \notag \\
&\quad+ \gamma_{sq}|A_\mathrm{q}(z,t_\mathrm{s} - \beta_{1s}(L-z))|^2 \notag\\
&\quad+ \gamma_{ip}|A_\mathrm{p}(z,t_\mathrm{i} - \beta_{1i}(L-z))|^2 \notag\\
&\quad+ \gamma_{iq}|A_\mathrm{q}(z,t_\mathrm{i} - \beta_{1i}(L-z))|^2  \Big ] \mathcal{A}(z,t_\mathrm{s},t_\mathrm{i}).
\end{align}
From this equation, a number of solutions can be obtained. However, as is the case for the nonlinear Schrödinger equation, it is unlikely that solutions can be found when both NPM and GVD are included. By neglecting GVD, the evolution equation turns into a simple first-order differential equation of the form
\begin{align}
\frac{\partial \mathcal{A}(z,t_\mathrm{s},t_\mathrm{i})}{\partial z} &= f(z,t_s, t_i) \mathcal{A}(z,t_\mathrm{s},t_\mathrm{i}) + g(z,t_s,t_i),\\
\mathcal{A}(0,t_\mathrm{s},t_\mathrm{i}) &= 0,
\end{align}
for which the solution, evaluated at $z = L$, is
\begin{equation}
\mathcal{A}(t_\mathrm{s},t_\mathrm{i}) = \int_0^L \ud z \, g(z,t_s, t_i) \exp\left (\int_{z}^L \ud z' f(z',t_s,t_i) \right ),
\end{equation}
which, due to the delta-function in $g(z,t_s,t_i)$ reduces to
\begin{align}
\JT(t_s, t_i) &= \frac{i\gamma}{\beta_{1s} - \beta_{1i}} A_p(z_c, t_c) A_q(z_c, t_c) \exp(i\theta_{\mathrm{NPM}})\notag \\
&\quad\times \exp\left (i\int_{z_c}^{L} \ud z' \Delta\beta_0(z') \right ) \Theta(z_c) \Theta(L-z_c), 
\end{align}
where $\Theta$ is the Heaviside function and the collision coordinates are defined as
\begin{align}
z_c &= L - \frac{t_s - t_i}{\beta_{1s} - \beta_{1i}}, &  t_c &= \frac{\beta_{1s} t_i - \beta_{1i}t_s}{\beta_{1s} - \beta_{1i}},
\end{align}
which can be interpreted as the creation point and time of the photon pair (such that the delta-function argument is zero when $z = z_c$). The nonlinear phase is thus
\begin{align}
\theta_\mathrm{NPM} = 2i \int_{z_c}^L \ud z \, &\Big [\gamma_{sp}|A_\mathrm{p}(z,t_\mathrm{s} - \beta_{1s}(L-z))|^2 \notag \\
&+ \gamma_{sq}|A_\mathrm{q}(z,t_\mathrm{s} - \beta_{1s}(L-z))|^2 \notag\\
&+ \gamma_{ip}|A_\mathrm{p}(z,t_\mathrm{i} - \beta_{1i}(L-z))|^2 \notag \\
&+ \gamma_{iq}|A_\mathrm{q}(z,t_\mathrm{i} - \beta_{1i}(L-z))|^2  \Big ],
\end{align}
which can be simplified if the pumps and their evolution are specified~\cite{bell2015effects, christensen2016temporally}.

\subsubsection{Solution with higher-order dispersion}

If instead of neglecting GVD, we neglect NPM and dispersion fluctuations, we can transform the evolution equation \eqref{evolution_equation} to the spectral domain:
\begin{align}
&\frac{\partial \mathcal{A}(z,\omega_\mathrm{s},\omega_\mathrm{i})}{\partial z} = \frac{i\gamma}{2\pi} \int \ud \omega A_\mathrm{p}(z,\omega_\mathrm{s} + \omega) A_\mathrm{q}(z,\omega_\mathrm{i} - \omega) \notag\\
&\qquad+ i\left [ \beta_{1\mathrm{s}} \omega_s + \beta_{1\mathrm{i}} \omega_i + \frac{1}{2}\beta_{2\mathrm{s}}\omega_s^2 + \frac{1}{2}\beta_{2\mathrm{i}}\omega_i^2 \right ] \mathcal{A}(z,t_\mathrm{s},t_\mathrm{i}),
\end{align}
where we used the Fourier transform convention $f(\omega) = \int \ud t f(t) \exp(i\omega t)$. This is again just a simple first-order differential equation, but the convolution (instead of the delta function in the previous section) makes a closed-form solution difficult. However, if a degenerate Gaussian pump with the initial amplitude $A_p(0,t) =  \sqrt{P_p}\exp(-\sigma_p^2 t^2/2)$, is assumed, an approximate solution can be found~\cite{koefoed2019effects}
\begin{align}
\JS(\omega_s, \omega_i) &= i\sqrt{\pi} \gamma L P_p \sigma_p^{-1} \exp \left (-\frac{(\omega_s + \omega_i)^2}{4\sigma_p^2} \right )  \\
&\quad \times \sinc\Bigg ( \Bigg [\frac{1}{4} \beta_{2p}(\omega_s + \omega_i)^2 -\beta_{1s}\omega_s - \frac{1}{2}\beta_{2s}\omega_s^2 \notag \\
&\qquad \qquad \qquad - \beta_{1i}\omega_i - \frac{1}{2}\beta_{2i}\omega_i^2 + \frac{\beta_{2p} \sigma_p^2}{2} \Bigg ] \frac{L}{2} \Bigg ), \notag
\end{align}
where $\beta_{2p}$ is the GVD experienced by the pump, $\sigma_p$ is the pump spectral width and $P_p$ is the pump power.

\section{Comparison of three effects in two schemes}

To illustrate the impact of each of the effects discussed in the previous section on the photon-pair state, we can consider each of them separately. We consider the impact on two-photon states with very low spectral correlations prior to introducing each effect. The amount of correlation is quantified by the post-heralding quantum purity of the remaining photon. The purity $0 \leq P \leq 1$ is calculated from a Schmidt decomposition of the joint amplitude~\cite{uren2005, bell2015effects} with a completely uncorrelated joint state leading to unity purity of the heralded photon.

We here consider two experimentally interesting examples of photon-pair-generation schemes using FWM that generate heralded photons of high quantum purity. The first is often referred to as asymmetric group-velocity matching, relying on one of the quantum fields being group-velocity matched to a degenerate Gaussian pump~\cite{garay2007photon}. Due the requirements on the group velocities this scheme has been realized with four-wave mixing in microstructured fibers where the dispersion can be carefully controlled~\cite{cohen2009tailored,halder2009nonclassical}. 

The second scheme, which we refer to as the collision scheme, relies on two non-degenerate pumps with identical Gaussian envelopes, but different group velocities, making a full temporal collision inside the waveguide. There have been suggestions to achieve this difference in pump speeds using chromatic dispersion~\cite{fang2013state}, waveguide birefringence~\cite{christensen2016temporally} and higher-order waveguide modes~\cite{koefoed2017spectrally}. We here focus on the special case where each quantum field is group-velocity matched to one of the pumps, e.g. $\beta_{1s} = \beta_{1p}$ and $\beta_{1i} = \beta_{1q}$. This case has been shown to be robust to NPM~\cite{christensen2016temporally} and HOD~\cite{koefoed2019effects}, but has yet to be experimentally demonstrated. For any given waveguide length, the pulses are timed so maximal overlap occurs at the waveguide midpoint. In the absence of disruptive effects, both schemes can achieve arbitrarily high single-photon purity as the waveguide length is increased.

The Gaussian pumps used in the two schemes take the form $A_p(z,t) =  \sqrt{P_p}\exp(-\sigma_p^2 t^2/2)$ so we can use all the analytical solutions from the previous section. We use the pulse duration $T_p = \sigma_p^{-1} = 1 \, \mathrm{ps}$, a difference between all non-copropagating fields of $\Delta\beta_1 = 1\times 10^{-11} \, \mathrm{s/m}$, a photon-pair generation probability of $R = 0.2$, GVD for all fields corresponding to $\beta_2 = 50 \times 10^{-26} \, \mathrm{s^2/m}$ and a waveguide length of $10 \, \mathrm{m}$ for the asymmetric scheme and $1 \, \mathrm{m}$ for the collision scheme, which is enough for a complete collision. These values are representative of a step-index silica fiber, but the magnitudes of the parameters can vary greatly between different platforms, waveguide types, wavelengths and other system parameters.

The dispersion fluctuations are modelled using a Brownian-motion model~\cite{koefoed2017spectrally} for the change in phase-matching frequency $\Delta\omega$ with $\Delta\beta_0(z) = \Delta\beta_1 \Delta\omega(z)$ and a standard deviation $\sigma_{\Delta\omega} = 0.5 \sigma_p$ with a correlation length of $10 \, \mathrm{cm}$. 

\begin{figure}[!htb]
    \centering
    \def\svgwidth{0.95\linewidth}
    
    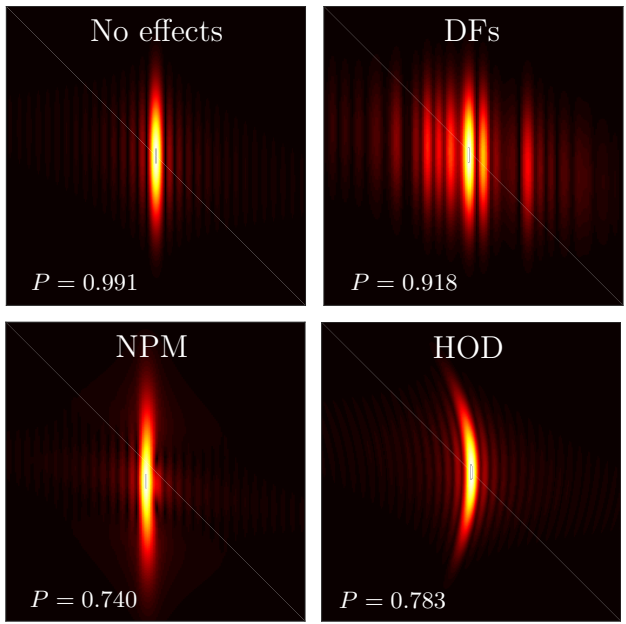
    \caption{Effect of dispersion fluctuations, nonlinear phase modulation and group-velocity dispersion on the joint spectral amplitude in the asymmetric scheme compared to no effects. The heralded photon purities are indicated in each case.}
    \label{fig:foureffects_asym}
\end{figure}

For the asymmetric scheme, the resulting two-photon state with each effect included is shown in Fig. \ref{fig:foureffects_asym}.
As an indication of the correlations introduced by each effect, the quantum purity~\cite{uren2005} which sets an upper limit on two-photon interference visibility, of the heralded photon is given in each case. In the asymmetric scheme, low spectral correlation and hence high post-heralding purity, comes from the narrow spectral distribution in one of the frequencies. Due to the large waveguide length, the phase-matching window is very narrow, leading to the state with no effects being highly uncorrelated and showing a purity of $P = 99.1 \, \%$. Dispersion fluctuations smears out the state in the diagonal direction. Even though there is significant distortion to the state, the purity is still high at $P = 91.8 \, \%$. This is because each vertical peak is still highly uncorrelated. Nonlinear phase modulation spectrally broadens the state and introduces phase correlations, reducing the purity for high generation rates to $P = 74.0 \, \%$. The impact of HOD is independent of generation rate, but depends strongly on pump duration. The effect of GVD is to introduce curvature to the state, reducing purity to $P = 78.3 \, \%$ for the dispersion chosen for this example.

\begin{figure}[!htb]
    \centering
    \def\svgwidth{0.95\linewidth}
    
    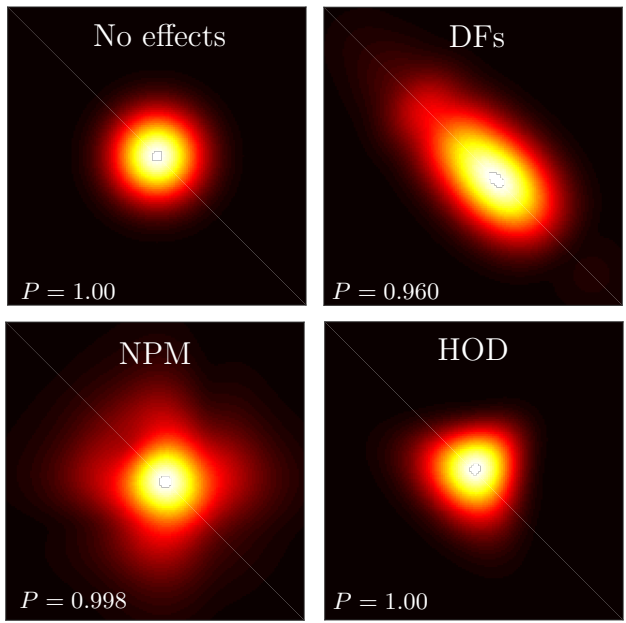
    \caption{Effect of dispersion fluctuations, nonlinear phase modulation and group-velocity dispersion on the joint spectral amplitude in the collision scheme compared to no effects. The heralded photon purities are indicated in each case.}
    \label{fig:foureffects_coll}
\end{figure}

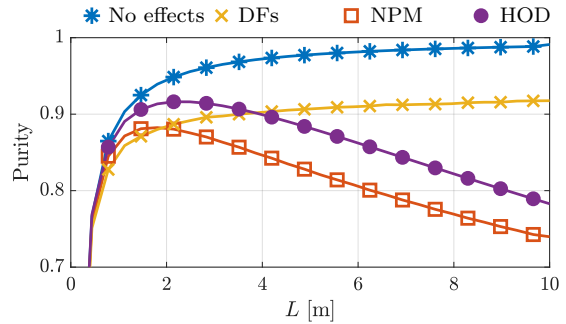
\begin{figure}[!htb]
    \centering
    \def\svgwidth{0.95\linewidth}
    
\begingroup%
  \makeatletter%
  \providecommand\color[2][]{%
    \errmessage{(Inkscape) Color is used for the text in Inkscape, but the package 'color.sty' is not loaded}%
    \renewcommand\color[2][]{}%
  }%
  \providecommand\transparent[1]{%
    \errmessage{(Inkscape) Transparency is used (non-zero) for the text in Inkscape, but the package 'transparent.sty' is not loaded}%
    \renewcommand\transparent[1]{}%
  }%
  \providecommand\rotatebox[2]{#2}%
  \newcommand*\fsize{\dimexpr\f@size pt\relax}%
  \newcommand*\lineheight[1]{\fontsize{\fsize}{#1\fsize}\selectfont}%
  \ifx\svgwidth\undefined%
    \setlength{\unitlength}{450bp}%
    \ifx\svgscale\undefined%
      \relax%
    \else%
      \setlength{\unitlength}{\unitlength * \real{\svgscale}}%
    \fi%
  \else%
    \setlength{\unitlength}{\svgwidth}%
  \fi%
  \global\let\svgwidth\undefined%
  \global\let\svgscale\undefined%
  \makeatother%
  \begin{picture}(1,0.51227552)%
    \lineheight{1}%
    \setlength\tabcolsep{0pt}%
    \put(0,0){\includegraphics[width=\unitlength,page=1]{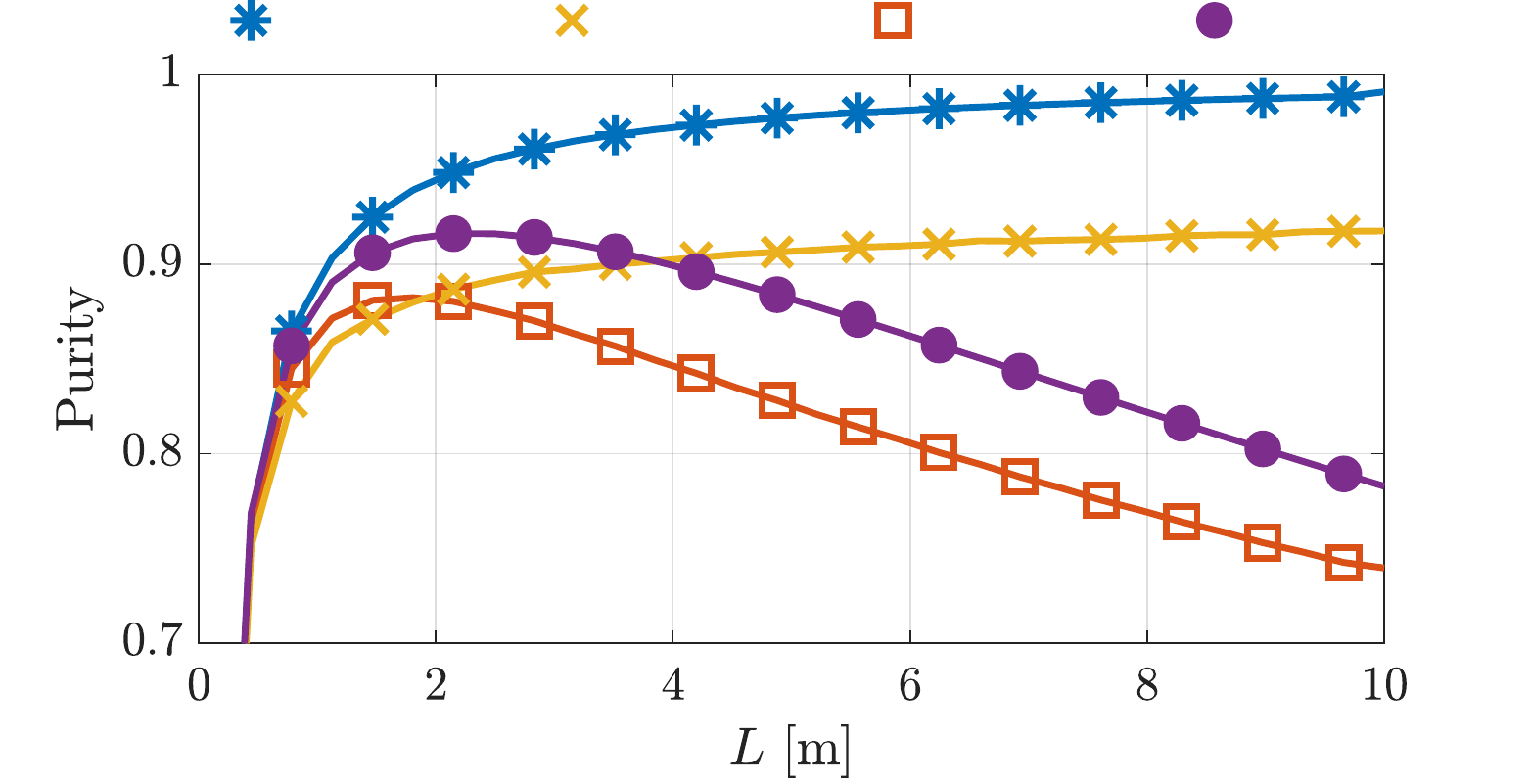}}%
    \put(0.19081285,0.48900074){\color[rgb]{0,0,0}\makebox(0,0)[lt]{\lineheight{1.25}\smash{\begin{tabular}[t]{l}\footnotesize No effects\end{tabular}}}}%
    \put(0.40115138,0.48941089){\color[rgb]{0,0,0}\makebox(0,0)[lt]{\lineheight{1.25}\smash{\begin{tabular}[t]{l}\footnotesize DFs\end{tabular}}}}%
    \put(0.61490138,0.48922209){\color[rgb]{0,0,0}\makebox(0,0)[lt]{\lineheight{1.25}\smash{\begin{tabular}[t]{l}\footnotesize NPM\end{tabular}}}}%
    \put(0.82419175,0.48923511){\color[rgb]{0,0,0}\makebox(0,0)[lt]{\lineheight{1.25}\smash{\begin{tabular}[t]{l}\footnotesize HOD\end{tabular}}}}%
  \end{picture}%
\endgroup%

    \caption{Heralded purity versus propagation length for the asymmetric scheme under no effects, dispersion fluctuations, nonlinear phase modulation and higher-order dispersion.}
    \label{fig:graph_asym}
\end{figure}

\begin{figure}[!htb]
    \centering
    \def\svgwidth{0.95\linewidth}
    
\begingroup%
  \makeatletter%
  \providecommand\color[2][]{%
    \errmessage{(Inkscape) Color is used for the text in Inkscape, but the package 'color.sty' is not loaded}%
    \renewcommand\color[2][]{}%
  }%
  \providecommand\transparent[1]{%
    \errmessage{(Inkscape) Transparency is used (non-zero) for the text in Inkscape, but the package 'transparent.sty' is not loaded}%
    \renewcommand\transparent[1]{}%
  }%
  \providecommand\rotatebox[2]{#2}%
  \newcommand*\fsize{\dimexpr\f@size pt\relax}%
  \newcommand*\lineheight[1]{\fontsize{\fsize}{#1\fsize}\selectfont}%
  \ifx\svgwidth\undefined%
    \setlength{\unitlength}{450bp}%
    \ifx\svgscale\undefined%
      \relax%
    \else%
      \setlength{\unitlength}{\unitlength * \real{\svgscale}}%
    \fi%
  \else%
    \setlength{\unitlength}{\svgwidth}%
  \fi%
  \global\let\svgwidth\undefined%
  \global\let\svgscale\undefined%
  \makeatother%
  \begin{picture}(1,0.51000001)%
    \lineheight{1}%
    \setlength\tabcolsep{0pt}%
    \put(0,0){\includegraphics[width=\unitlength,page=1]{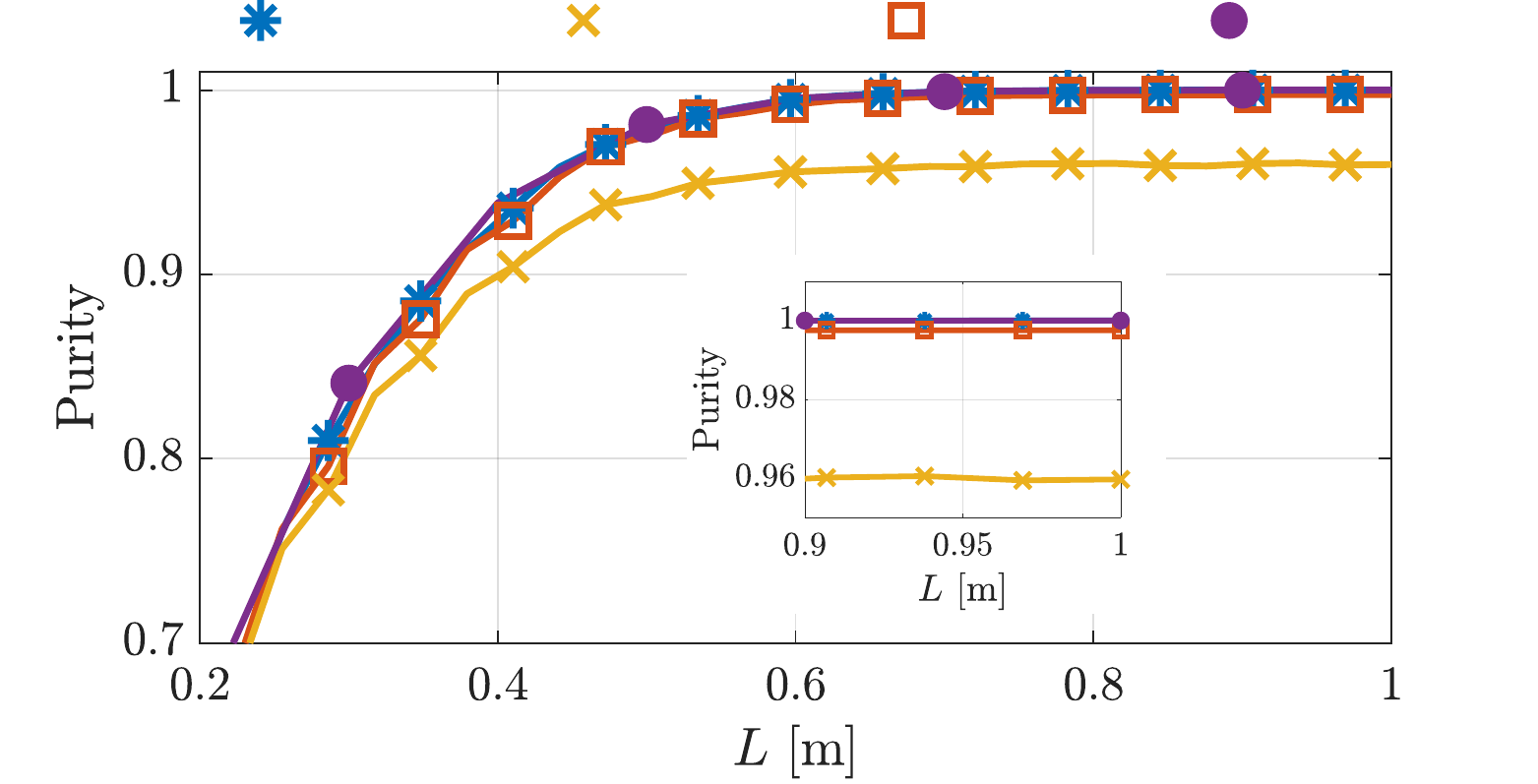}}%
    \put(0.19624797,0.48672523){\color[rgb]{0,0,0}\makebox(0,0)[lt]{\lineheight{1.25}\smash{\begin{tabular}[t]{l}\footnotesize No effects\end{tabular}}}}%
    \put(0.40658649,0.48713538){\color[rgb]{0,0,0}\makebox(0,0)[lt]{\lineheight{1.25}\smash{\begin{tabular}[t]{l}\footnotesize DFs\end{tabular}}}}%
    \put(0.62033649,0.48694658){\color[rgb]{0,0,0}\makebox(0,0)[lt]{\lineheight{1.25}\smash{\begin{tabular}[t]{l}\footnotesize NPM\end{tabular}}}}%
    \put(0.82962685,0.4869596){\color[rgb]{0,0,0}\makebox(0,0)[lt]{\lineheight{1.25}\smash{\begin{tabular}[t]{l}\footnotesize HOD\end{tabular}}}}%
    \put(0,0){\includegraphics[width=\unitlength,page=2]{graph_coll.pdf}}%
  \end{picture}%
\endgroup%

    \caption{Heralded purity versus propagation length for the collision scheme under no effects, dispersion fluctuations, nonlinear phase modulation and higher-order dispersion. Inset shows magnified view of the indicated region.}
    \label{fig:graph_coll}
\end{figure}

As Fig. \ref{fig:foureffects_asym} suggests, the asymmetric scheme is vulnerable to parasitic effects due to its narrow spectral distribution in either the signal or idler direction. The same three effects in the collision scheme is shown in Fig. \ref{fig:foureffects_coll}.
We see similar patterns of smearing, broadening and distortion from DFs, NPM and HOD, respectively, as for the asymmetric scheme. However, as suggested by earlier research, this scheme is overall much less susceptible to degradation in purity due to these effects. To compare these two schemes quantitatively under each of these effects, we calculate the purity as a function of propagation length for the two schemes. Note, that in the case of the collision scheme, a shorter length means an incomplete pulse collision. The HOD calculation for the collision scheme is carried out using the numerical procedure outlined in this paper since no analytical solution has been discovered. The results for the asymmetric scheme is shown in Fig. \ref{fig:graph_asym}.
The monotonic increase in purity is broken by the introduction of both NPM and HOD to the system. In each case, the introduced effects creates a limit on the achievable purity and leads to an optimal propagation length, which is important to be aware of in experimental designs. In the case of DFs, the purity does not display the same behavior. In this case, it still increases, but at a much slower rate. In addition, even though the achievable purity may be high, even in the presence of DFs, the achievable two-photon-interference visibility between distinct sources with DFs may be low~\cite{koefoed2017spectrally}. The corresponding graph for the collision scheme is shown in Fig. \ref{fig:graph_coll}.
As expected, this scheme is much more robust to degradation in purity due to the three effects. In all cases, a complete collision ($L \gtrsim 0.6 \, \mathrm{m}$) is ideal with only DFs showing a significant effect, even for these values for the effect parameters. Previous research has shown that, in some cases, fiber dispersion can be designed to be robust to such fluctuations~\cite{koefoed2017spectrally}.

\FloatBarrier

\section{Conclusion}

We have developed a general Schrödinger-picture framework to describe the evolution of the joint amplitude in photon-pair generation by four-wave mixing. This framework allows for the inclusion of effects, such as longitudinal dispersion fluctuations, nonlinear phase modulation from the classical pumps and higher-order dispersion. We described a numerical split-step scheme to solve the general propagation problem and gave a number of special-case analytical solutions. Finally, we used the analytical and numerical solutions to compare two experimentally interesting schemes, the asymmetric scheme and the collision scheme, for generating quantum-mechanically pure heralded photons. We found that the asymmetric scheme is sensitive to all three parasitic effects considered here, while the collision scheme is robust to all three. This makes the collision scheme interesting from an experimental point of view, since very high purities could be achievable in real system with considerably less effort.

\section*{Acknowledgement}

This work was supported by the Danish Council for
Independent Research (DFF) (4184-00433).

\bibliography{../library}

\end{document}